\documentclass[prl,aps,showpacs,twocolumn,preprintnumbers,amsmath,amssymb,superscriptaddress,longbibliography,nofootinbib]{revtex4-2}
\usepackage[english]{babel}

\usepackage{graphicx}
\usepackage[colorlinks=True,linkcolor=blue,citecolor=blue,urlcolor=blue]{hyperref}
\usepackage{xcolor}
\usepackage{bm}

\begin{document}

\title{Stacking-tuned superconductivity and competing charge-density-wave states in NbSe$_2$}

\author{Sandra Sajan}
\email{These authors contributed equally to this work.}
\affiliation{Donostia International Physics Center, Donostia-San Sebasti\'an, Spain}

\author{Xinze Yang}
\email{These authors contributed equally to this work.}
\affiliation{Department of Physics, Yale University, New Haven, CT, USA}
\affiliation{Energy Sciences Institute, Yale University, West Haven, CT, USA}

\author{Haojie Guo}
\affiliation{Donostia International Physics Center, Donostia-San Sebasti\'an, Spain}

\author{Tarushi Agarwal}
\affiliation{Department of Physics, Indian Institute of Science Education and Research Bhopal, Bhopal, India}

\author{Samuel Ma\~nas-Valero}
\affiliation{Institute of Molecular Science (ICMOL), University of Valencia, Paterna-Valencia, Spain}

\author{Carla Boix-Constant}
\affiliation{Institute of Molecular Science (ICMOL), University of Valencia, Paterna-Valencia, Spain}
\affiliation{Department of Materials Science \& Metallurgy, University of Cambridge, Cambridge, United Kingdom}

\author{Eugenio Coronado}
\affiliation{Institute of Molecular Science (ICMOL), University of Valencia, Paterna-Valencia, Spain}

\author{Fernando de Juan}
\affiliation{Donostia International Physics Center, Donostia-San Sebasti\'an, Spain}
\affiliation{Ikerbasque, Basque Foundation for Science, Bilbao, Spain}

\author{Eduardo~H.~da~Silva~Neto}
\affiliation{Department of Physics, Yale University, New Haven, CT, USA}
\affiliation{Energy Sciences Institute, Yale University, West Haven, CT, USA}
\affiliation{Department of Applied Physics, Yale University, New Haven, CT, USA}

\author{Ravi P. Singh}
\affiliation{Department of Physics, Indian Institute of Science Education and Research Bhopal, Bhopal, India}

\author{Maria N. Gastiasoro}
\email{Corresponding author: maria.ngastiasoro@dipc.org}
\affiliation{Donostia International Physics Center, Donostia-San Sebasti\'an, Spain}

\author{Miguel M. Ugeda}
\email{Corresponding author: mmugeda@dipc.org}
\affiliation{Donostia International Physics Center, Donostia-San Sebasti\'an, Spain}
\affiliation{Ikerbasque, Basque Foundation for Science, Bilbao, Spain}
\affiliation{Centro de F\'isica de Materiales CSIC-UPV/EHU, San Sebasti\'an, Spain}

\date{\today}
\begin{abstract}

Layer stacking provides a powerful yet underexplored route for reshaping collective quantum order in van der Waals materials. Here we use high-resolution scanning tunneling microscopy and spectroscopy to show that the stacking sequence alone can qualitatively transform the charge density and superconducting orders in NbSe$_2$, while preserving the same in-plane atomic structure. Comparing the 4Ha and 2H polytypes, we find that, unlike the ubiquitous triangular incommensurate $3Q^\mathrm{I}$ order of 2H-NbSe$_2$, 4Ha-NbSe$_2$ hosts two competing CDW states with no measurable correlation with local strain: a unidirectional commensurate $1Q^\mathrm{C}$ phase and a triangular incommensurate $3Q^\mathrm{I}$ phase, with $Q^\mathrm{I}=Q^\mathrm{C}+\delta$. We introduce a phase-resolved analysis that directly maps the gradient of the CDW phases and reveals vortices bound to the $1Q^\mathrm{C}$–$3Q^\mathrm{I}$ phase boundaries. These vortices accommodate the momentum mismatch $\delta$ through abrupt $2\pi$ phase slips, providing a  mechanism by which distinct charge orders coexist. Superconductivity is also reshaped by stacking, while both polytypes exhibit multiband pairing.

\end{abstract}

\maketitle


In metallic transition metal dichalcogenides (TMDs), layer stacking defines symmetry and governs the electronic structure. It modifies interlayer coupling, hence interlayer hybridization and the local symmetry environment~\cite{Wilson1975,Xi2016,Heil2017,Yokoya2001}. Stacking can therefore reshape the band structure and generate spin-orbit-driven effects, including splitting of spin-degenerate bands and spin-texture formation, without altering the chemical composition~\cite{Ribak2020Chiral,Wang2022NbS2,Butler2020,Gao2020Magnetoresistance,Law2017,prx2020stacking,natphys2021tmdspin}. Consequently, it can govern collective phases such as charge-density waves (CDWs), superconductivity, correlated states, and magnetic order~\cite{Wang2022NbS2,Ribak2020Chiral,Gao2020Magnetoresistance,ManasValero2021,Wang2020Mott,Sipos2008,Law2017,He2018Review,Butler2020,Ma2016,Martino2026}. However, disentangling the role of crystal symmetry and interlayer coupling, as encoded in the stacking sequence, from other intertwined effects on collective phases remains challenging. 

\begin{figure}[]
    \centering
    \includegraphics[width=\columnwidth]{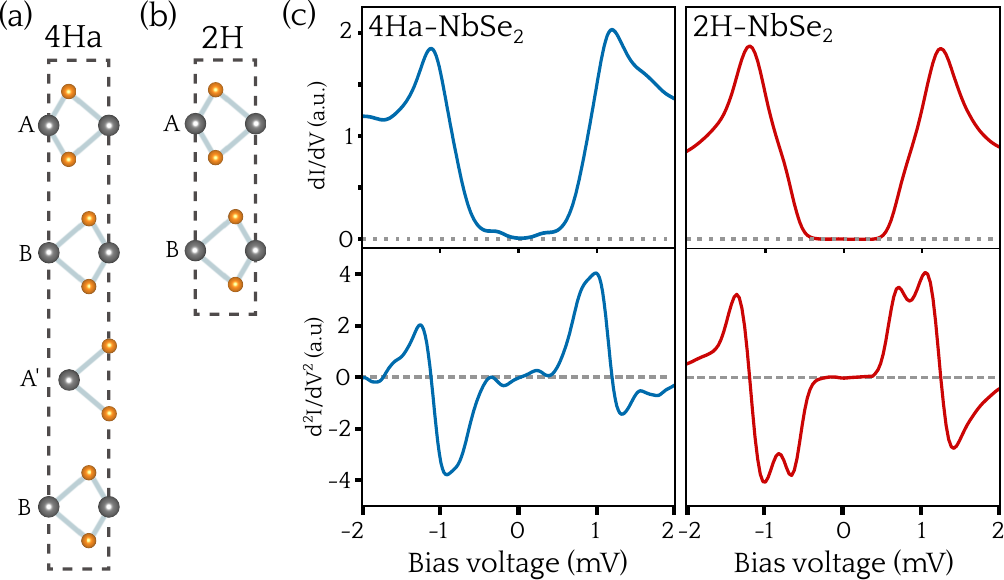}
    \caption{(a,b)  Unit cell of 4Ha-$NbSe_{2}$ and 2H-$NbSe_{2}$ respectively, (c) Representative $dI/dV$ and corresponding derivatives ($d^2I/dV^2$) of the 4Ha and 2H polytypes ($V_{\mathrm{ac}} = 20\,\mu\mathrm{V},\, $T = 0.34 K).}
    \label{Fig1}
\end{figure}

NbSe$_2$ provides an ideal platform for this purpose, as its stable 2H and 4Ha polytypes both host CDW order and superconductivity~\cite{Kadijk1971,Zhou2023,Naik2018NbSe2,Naik2011NbSe2,Shin2005Te2,Kikuchi1998NbSe2}. Figures~\ref{Fig1}(a) and~\ref{Fig1}(b) show the crystal structures of 4Ha- and 2H-NbSe$_2$, highlighting their respective structural units. In 2H-NbSe$_2$, trigonal-prismatic NbSe$_2$ layers stack in an alternating AB sequence along the $c$ axis, giving rise to the centrosymmetric, non-symmorphic space group $P6_3/mmc$. In contrast, 4Ha-NbSe$_2$ exhibits a modified periodic stacking sequence: after two 2H-like layers, the third layer is translated within the $ab$ plane by $(2\mathbf{a}/3+\mathbf{b}/3)$ relative to the second layer, requiring four layers to complete the unit cell (ABA'B'). This stacking modification lowers the crystal symmetry to the noncentrosymmetric, symmorphic space group $P\bar{6}m2$.

How the distinct stacking sequences of 2H- and 4Ha-NbSe$_2$ affect their electronically ordered phases remains poorly understood, particularly at the nanoscale. Experimentally, 4Ha-NbSe$_2$ has received limited attention: while superconducting ($T_c \sim 6.4$ K) and charge-density-wave ($T_{\mathrm{CDW}} \sim 40$ K) transitions are established, a microscopic understanding of its CDW structure and interplay with superconductivity remains incomplete, unlike in the extensively studied 2H phase. Theoretically, inversion-symmetry breaking in 4Ha-NbSe$_2$ provides a structural basis for distinct electronic behavior compared with centrosymmetric 2H-NbSe$_2$, including spin-split bands and modified interlayer coupling~\cite{Zhou2023}. Consistently, \emph{ab initio} studies show substantial splitting of Nb-derived bands, directly linking stacking to the electronic structure and enabling Ising protection of Cooper pairs~\cite{Patra2025,Volavka, Ni2026}.

Motivated by this contrast, we performed a systematic STM/STS comparison of 4Ha- and 2H-NbSe$_2$. High resolution STS reveals that while superconductivity shows multiband features in both polytypes, the 4Ha stacking reduces the superconducting energy scales relative to the 2H phase. The most pronounced changes occur in the CDW order. Instead of the triangular incommensurate $3Q^\mathrm{I}$ order found in 2H-NbSe$_2$, 4Ha-NbSe$_2$ displays an intrinsic coexistence of two distinct charge-ordered states: a unidirectional commensurate $1Q^\mathrm{C}$ phase and a triangular incommensurate $3Q^\mathrm{I}$ phase, with $Q^\mathrm{I}=Q^\mathrm{C}+\delta$. By mapping the local CDW phase gradients, we identify vortices pinned to the boundaries between these two phases. These vortices act as localized $2\pi$ phase slips that absorb the momentum mismatch $\delta$, allowing the commensurate and incommensurate charge orders to coexist at the nanoscale.

{ \sl Results} ---
Single crystals of 4Ha-NbSe$_2$ and 2H-NbSe$_2$ were grown by chemical vapor transport (CVT), exfoliated under ultrahigh vacuum, and transferred \textit{in situ} into a cryogenic STM (see End Matter). To assess the effect of stacking on superconductivity, we compare high-resolution $dI/dV$ spectra acquired on both polytypes. In 4Ha-NbSe$_2$, the spectra show a two-step gap structure, with coherence peaks near $\pm1.1~\mathrm{mV}$ and weak low-energy shoulders near $\pm0.3~\mathrm{mV}$. Fits using a self-consistent anisotropic two-band McMillan model (Sec.~S2) yield $\Delta_1 = 0.987\pm0.082 ~\mathrm{meV}$ and $\Delta_2 = 0.269\pm0.130~\mathrm{meV}$, with sizable quasiparticle broadening and a normalized tunneling weight $w \approx 0.98$ (definition in Sec.~S2) for the larger-gap component (see Sec.~S3 for magnetic field dependence). These values agree well with previous indirect estimates from specific heat~\cite{Zhou2023,Zhou2025}, lower-critical-field/penetration-depth analyses~\cite{Zhou2025}, and $\mu$SR measurements~\cite{vonRohr2019,Islam2025}. In contrast, 2H-NbSe$_2$ exhibits sharper coherence peaks and a clearly resolved secondary gap, with $\Delta_1 \approx1.08~\mathrm{meV}$ and $\Delta_2 \approx 0.72~\mathrm{meV}$, reduced broadening, and a more balanced band contribution ($w \approx 0.68$). Thus, although both polytypes show multiband superconductivity, they differ markedly in the spectral weight and visibility of the smaller gap.

\begin{figure*}[]
    \includegraphics[width=\textwidth]{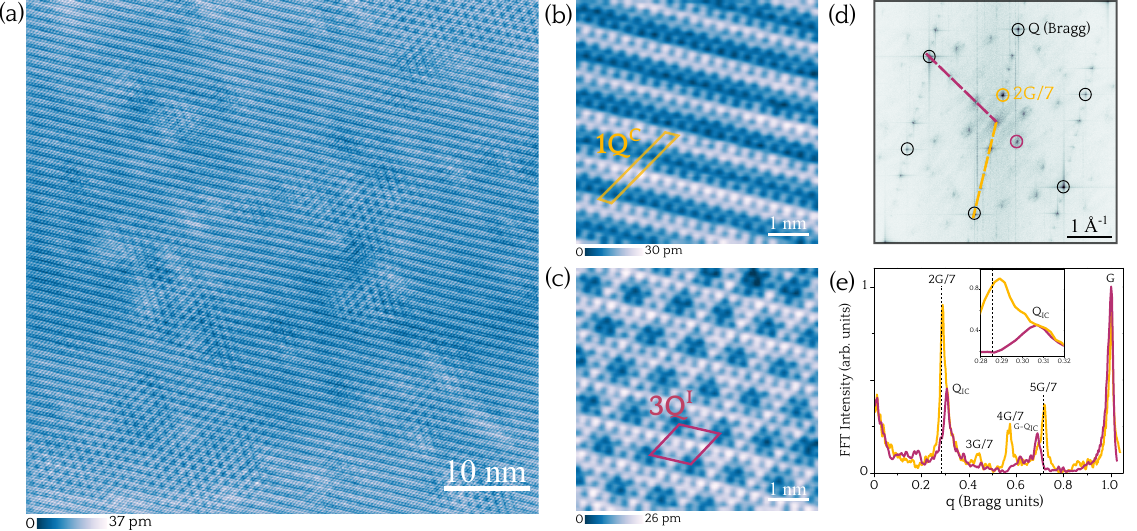}
        \caption{(a) Large-scale STM topograph of 4Ha-NbSe$_2$ showing coexisting unidirectional commensurate $1Q^\mathrm{C}$ and triangular incommensurate $3Q^\mathrm{I}$ CDW phases ($V_s = -0.098$~V, $T = 0.34$~K). (b),(c) Zoomed-in STM images of the $1Q^\mathrm{C}$ and $3Q^\mathrm{I}$ phases, respectively ($V_s = -0.1$~V, $T = 0.34$~K). (d) FFT of panel (a), showing the $1Q^\mathrm{C}$ feature along one crystallographic direction and $3Q^\mathrm{I}$ features along all three symmetry-equivalent directions. (e) FFT intensity profiles along the $1Q^\mathrm{C}$ and $3Q^\mathrm{I}$ directions marked in panel (d).}
        \label{fig2}
\end{figure*}

Given the recently reported twofold upper critical field anisotropy in the superconducting state \cite{Ni2026}, despite the underlying threefold crystal symmetry, it is natural to ask whether signatures of rotational symmetry-breaking are also manifested in the charge-ordered state. To address this question, we next investigate the CDW order in 4Ha-NbSe$_2$ using large-scale STM imaging. Figure~\ref{fig2}(a) shows a representative topography image of an atomically flat $55 \times 55$~nm$^2$ region, revealing the coexistence of unidirectional $1Q^\mathrm{C}$ and triangular $3Q^\mathrm{I}$ CDW phases. Importantly, this coexistence occurs on atomically flat terraces, without the height contrast or buckling typically associated with strain-induced stripe order in 2H-NbSe$_2$\cite{Soumyanarayanan2013,Gao2018}. A continuous evolution from the conventional $3Q^\mathrm{I}$ CDW superlattice to a unidirectional $1Q^\mathrm{C}$ modulation is observed, where the ordering wave vector aligns along one of the three symmetry-equivalent $3Q^\mathrm{I}$ directions. Zoomed-in images of the $1Q^\mathrm{C}$ and $3Q^\mathrm{I}$ regions are shown in Figs.~\ref{fig2}(b) and \ref{fig2}(c), respectively, where the CDW superlattices are visible. The corresponding fast Fourier transform (FFT) of Fig.~\ref{fig2}(a), displayed in Fig.~\ref{fig2}(d), reveals a CDW peak for the $1Q^\mathrm{C}$ phase along one crystallographic direction at $\frac{2}{7}\bm{G}_1$ (yellow line), whereas the $3Q^\mathrm{I}$ phase produces peaks along all three symmetry-related directions with $Q^\mathrm{I}_n=\frac{2}{7}G_n+\delta_n$ with an incommensuration $\delta_n\approx 0.021 G_n$ (magenta line). Figure~\ref{fig2}(e) shows FFT intensity profiles along the $1Q^\mathrm{C}$ direction (yellow in Fig.~\ref{fig2}(d)) and along one of the $3Q^\mathrm{I}$ directions (magenta in Fig.~\ref{fig2}(d)), highlighting the relative spectral weight of the two components.


\begin{figure*}[]
    \centering
\includegraphics[width=\linewidth]{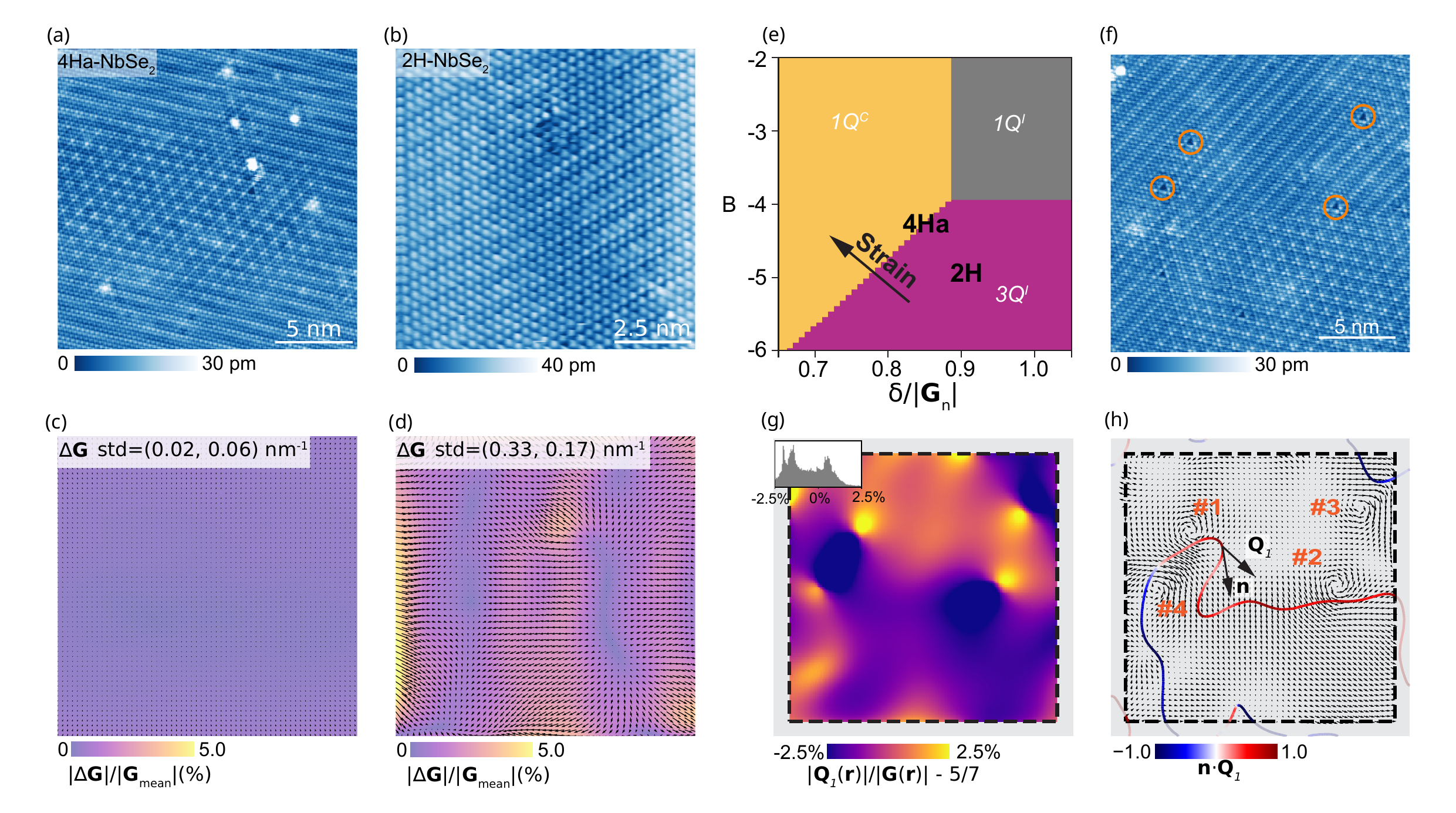}
    \caption{Topography of a region showing the coexistence of $1Q^\mathrm{C}$ and $3Q^\mathrm{I}$ CDW on (a) 4Ha and (b) 2H. (c) and (d) show the corresponding vector-field plots of the change in the Bragg vector relative to the mean Bragg vector in the same region. The background color map shows the percentage change in the magnitude of the Bragg vector. (e) GL phase diagram [Eq.\eqref{eq:GL}]. (f) STM topograph of 4Ha with $1Q^\mathrm{C}$ and $3Q^\mathrm{I}$ regions. (g) Corresponding commensurability map, $|\bm{\tilde{Q}}_1(\bm{r})|/|\bm{G}(\bm{r})| - 5/7$, with perfect commensuration at the origin. (h) Vector field of $\bm{\nabla}\varphi (\bm r)$ (black arrows) for the commensurate vector $\bm{\tilde{Q}}_1$. The boundary between $1Q^\mathrm{C}$ and $3Q^\mathrm{I}$ regions is depicted by solid lines, colored according to their local tangential orientation $\hat{\bm n}(\bm r)$ with respect to $\bm{Q}_1$, and the vortices are numbered $\#1$ through $\#4$. }
    \label{Fig3}
\end{figure*}
Unidirectional CDW order has previously been observed on the surface of 2H-NbSe$_2$ in regions of local strain \cite{Soumyanarayanan2013,Gao2018}. In contrast, our results indicate that the unidirectional $1Q^C$ CDW in 4Ha is not measurably strain-driven. As shown in Fig.\,\ref{Fig3}(a) and Fig.\,\ref{Fig3}(b), the phase separation between the $1Q^C$ and $3Q^I$ CDW on the 4Ha surface occurs on an otherwise uniformly flat surface, with no evidence of buckling. By comparison, in 2H-NbSe$_2$ the $1Q^C$ CDW correlates with dark-contrast regions, suggesting the presence of strain.

To further examine the relationship between strain and CDW state formation, we implemented a local wave-vector measurement algorithm based on wave-front detection, developed by some of the present authors (End Matter). Because strain distorts the lattice and thereby changes the local Bragg vector, the spatial variation of the Bragg vector provides a useful measure of strain strength. Using our algorithm, we measure a single Bragg vector locally and compute its deviation from the global average,
$\Delta\mathbf{G} = \mathbf{G}(\mathbf{r}) - \bar{\mathbf{G}}$.
The vector-field maps of $\Delta\mathbf{G}$ are shown in Fig.\,\ref{Fig3}(c) for 4Ha and Fig.\,\ref{Fig3}(d) for 2H. For the latter, the Bragg-vector fluctuations clearly track the transition between $1Q^C$ and $3Q^I$ domains. In contrast, within the sensitivity of our method, we do not observe a clear Bragg-vector fluctuation in the $1Q^C$–$3Q^I$ coexistence region of the 4Ha polytype.





We now use the Ginzburg-Landau (GL) formalism to understand the origin of the appearance of commensurate unidirectional $1Q^\mathrm{C}$ regions coexisting with incommensurate triangular $3Q^\mathrm{I}$ regions in 4Ha, without apparent strain or buckling in the samples. For this, we adapt the free energy considered in the past for NbSe$_2$ \cite{McMillan75,nakanishi1978,Flicker2015} to include the possibility of commensurate states with period 7 modulation. The free energy for a multi-component order parameter $\bm\phi=(\phi_1,\phi_2,\phi_3)$ with a $\phi_n$ component for each ordering vector $\bm{Q}_n\approx \frac{2}{7}\bm{G}_n$ takes the following form for a given layer,
\begin{align}
\label{eq:GL}
    &\mathcal{F}_\phi=\sum_n \left[ 
    A|\phi_n|^2+A_\delta|(\bm \delta_n+i \bm{\partial})\phi_n|^2
    +\frac{B}{3} \; {\rm Re}[\phi_1\phi_2\phi_3]\right.\nonumber\\
    &\left.+C|\phi_n|^2|\phi_{n+1}|^2 +D|\phi_n|^4+E \;{\rm Re}[\phi_n^7]+F|\phi_n|^8\right].
\end{align} 
The term with coefficient $E$ is the commensurability energy for a period seven modulation, which makes a commensurate CDW state more energetically favorable than an incommensurate one. The last term, with coefficient $F$, is required to obtain a stable free energy. The selection between $1Q^\mathrm{C}$ and $3Q^\mathrm{I}$ states is given by the cubic term, with coefficient $B$, and the quartic terms $C$ and $D$. 
In fact, as shown in Fig. \ref{Fig3}(e), this free energy has a first order transition from a $3Q^\mathrm{I}$ incommensurate (pink) to a $1Q^\mathrm{C}$ commensurate (orange) state as a function of either $\bm \delta_n$ or $B$. Near this transition, real space phase coexistence between these two phases is expected, as observed in 4Ha-NbSe$_2$ [Fig. \ref{fig2}(a)-(c)]. 
This situation is very different for the 2H polytype, which is deeper in the $3Q^\mathrm{I}$ incommensurate region of the phase diagram, with $1Q^\mathrm{C}$ states having been reported only under considerable strain \cite{Soumyanarayanan2013}, in agreement with our results in Fig. \ref{Fig3}(b) and (d). 

In the $1Q^\mathrm{C}$ regions of 4Ha, besides the main peaks at $\bm Q_1=\frac{2}{7}\bm G_1 (\frac{5}{7}\bm G_1)$, Fourier analysis shows additional weaker peaks at $\bm q_1=\frac{4}{7}\bm{G}_1 (\frac{3}{7}\bm{G}_1)$ in Fig. \ref{fig2}(e). Here, the wavevectors in parenthesis are the harmonics $\bm{\tilde{k}}=-\bm k+\bm G_1$ of an ordering wavevector $\bm k$. We thus propose the existence of a secondary multi-component order parameter, $\bm\eta=(\eta_1,\eta_2,\eta_3)$, with ordering vector $\bm{q}_n=2\bm{Q}_n=\frac{4}{7}\bm{G}_n$, and hence transforming as the square of the primary order parameter $\eta_n\sim \phi_n^2$. Then the term $\mathcal{F}_{\phi\eta}=\lambda\sum_n\left(\eta_n^*\right)^3\phi_n$ coupling both order parameters is also allowed in the GL theory.
This term is also a commensurability energy favoring $1Q^\mathrm{C}$ states, which explains the emergence of $\bm q_1=\frac{4}{7}\bm{G}_1 (\frac{3}{7}\bm{G}_1)$ peaks [Fig.\ref{fig2}(e)] in addition to the primary peaks $\bm Q_1=\frac{2}{7}\bm G_1 (\frac{5}{7}\bm G_1)$.

The interface of a $1Q^\mathrm{C}$ commensurate region and a $3Q^\mathrm{I}$ incommensurate region must accommodate their momentum difference $\bm\delta$, 
and the system does it with local phase slips, as we show in the following. 
To extract the phase information, we again rely on our local wave-vector measurement algorithm. For a plane wave with a slowly varying local phase fluctuation, we can expand
\begin{equation}
    e^{i\left[\bm k \cdot \bm{r}+\varphi(\bm{r})\right]}\approx e^{i\left[\bm k \cdot \bm{r}+\varphi_0+\nabla\varphi_{\bm{r}_0}(\bm{r}-\bm{r}_0)\right]}= e^{i\left[\left(\bm k +\nabla\varphi_{\bm{r}_0}\right)\cdot \bm{r}+...\right]}
\end{equation}
so that the change in the local wave vector is approximately equal to the phase gradient $\nabla\varphi(\bm{r})$. In this experiment, we track the phase variation of the $\tilde{\bm Q}_1=\frac{5}{7}\bm{G}_{1}$ harmonic peak. First, maps of the Bragg vector ($\tilde{\bm{G}}_{1}(\bm{r})$) and the CDW wave vector ($\tilde{\bm{Q}}_{1}(\bm{r})$) are obtained using the local wave vector measurement algorithm. Then, the nominal CDW vector is defined as $\tilde{\bm Q}_{1, \mathrm{avg}}=5/7\tilde{\bm{G}}_{1, \mathrm{avg}}$ where $\tilde{\bm{G}}_{1, \mathrm{avg}}$ is the averaged Bragg vector. Finally, the phase gradient is defined as the deviation of the local CDW wave vector from this nominal value, i.e. $\nabla\varphi(\bm{r}) = \tilde{\bm{Q}}_{1}(\bm{r}) - \tilde{\bm Q}_{1, \mathrm{avg}}$.
The vector-field plot of $\nabla\varphi(\mathbf{r})$, for the topograph in Fig.\,\ref{Fig3}(f), is shown in Fig.\,\ref{Fig3}(h), with a clear change in the phase gradient between the $1Q^\mathrm{C}$ and $3Q^\mathrm{I}$ regions. To further highlight this difference, we calculate the deviation of the local CDW wavevector from commensurability $|\tilde{\bm Q}_1(\bm r)| / |\tilde{\bm{G}}(\bm{r})| - 5/7$. 
Both the spatial map of $|\tilde{\bm Q}_1(\bm r)| / |\bm{G}(\bm{r})| - 5/7$ in Fig.\,\ref{Fig3}(g) and its two-peaked histogram (inset) demonstrate the incommensurability of the $3Q^\mathrm{I}$ region.

\begin{figure}[]
    \centering
    \includegraphics[width=\columnwidth]{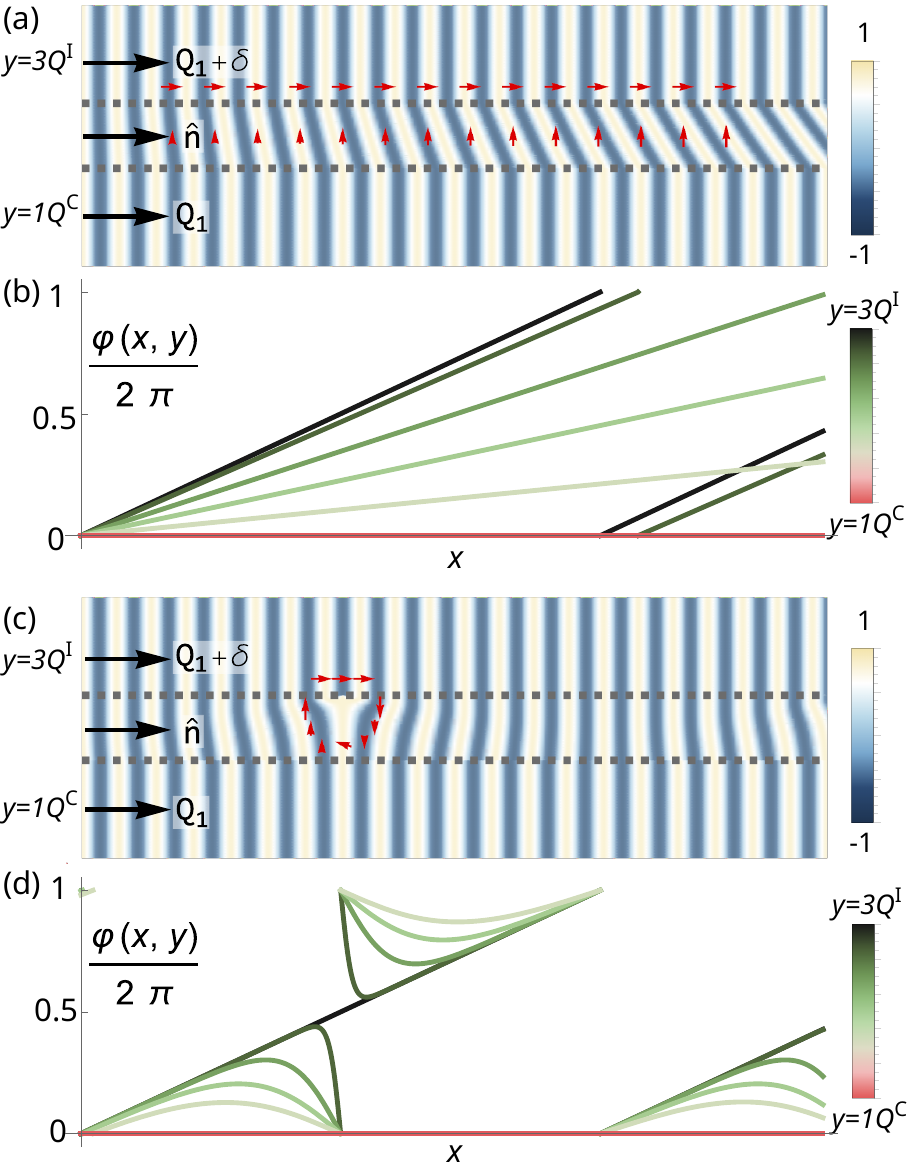}
    \caption{(a) Charge density at the phase boundary (region in between dashed gray lines) between commensurate region $1Q^\mathrm{C}$ and an incommensurate region $3Q^\mathrm{I}$ for the first component $\bm Q^\mathrm{I}_1=\bm Q_1+\bm \delta$. (b) The $3Q^\mathrm{I}$ state adapts its phase $\varphi(\bm{r})$ to match the $1Q^\mathrm{C}$ state along the boundary, gradually evolving from $\varphi_{3Q^\mathrm{I}}=\delta x$ to $\varphi_{1Q^\mathrm{C}}=0$. Its gradient (red arrows in (a)) shows no vortex. (c)-(d) Alternative way for the phase to adapt with an abrupt $2\pi$ slip, with a vortex appearing in its gradient. }
    \label{Fig4}
\end{figure}

Apart from the CDW incommensurability, vortices appear in $\bm{\nabla}\varphi(\bm r)$ at the boundary between the $1Q^\mathrm{C}$  and $3Q^\mathrm{I}$ regions (numbered $\#1$ through $\#4$ in Fig. \ref{Fig3}(h)), arising in the vicinity of defects, highlighted in Fig. \ref{Fig3}(f). Notably, these vortices are unlikely to be lattice dislocations since the corresponding vector-field map of the Bragg peak is vortex-free (See Supplementary Material). Strikingly, the chirality of the vortices follows the local boundary orientation, with counter-clockwise rotation ($\#1$ and $\#2$) for parallel boundaries ($\hat{\bm{n}}\cdot \hat{\bm{Q}}_1>0$), and clockwise rotation ($\#3$ and $\#4$) for antiparallel boundaries ($\hat{\bm{n}}\cdot \hat{\bm{Q}}_1<0$). 

The presence of each vortex signals local phase slips from its circulation, $\Delta\varphi=\oint_C \bm{\nabla}\varphi\cdot d\bm{l}=\pm 2\pi$, with the sign being determined by the chirality of the vortex. If the boundary is perpendicular to the ordering wavevector ($\hat{\bm{n}}\cdot \bm{Q}_1=0$), the ICDW can accommodate for the momentum difference between the $1Q^\mathrm{C}$ and $3Q^\mathrm{I}$ regions by a trivial phase slip $\Delta \varphi$ \emph{across} the boundary.  In the more interesting case of a finite parallel alignment component between the boundary and the ordering vector ($\hat{\bm n}\cdot \bm{Q}_1\neq 0$), an equivalent mechanism requires a phase change $\varphi(x)$ of the ICDW \emph{along} the boundary. 


This situation is illustrated in Fig. \ref{Fig4}(a), where the $3Q^\mathrm{I}$ ICDW (top), with its first component $\bm{Q}^\mathrm{IC}_1=(Q_1+\delta)\hat{\bm n}$ and corresponding $\rho_{3Q^\mathrm{I}}(\bm r)=\rho_0\cos\left( (Q_1+\delta) x+\theta_0(y)\right)\equiv\cos\left(Q_1 x+\varphi(x, y)\right)$, smoothly adapts its phase $\varphi(x, y)$ along the boundary, delimited by the gray dashed lines. As shown in  Fig. \ref{Fig4}(b), the phase evolves from $\varphi(x, y=3Q^\mathrm{I})=\delta x$ (dark green), gradually diminishing its slope until it reaches the other side of the boundary with $\varphi(x, y=1Q^\mathrm{C})=0$ (pink), to match the $1Q^\mathrm{C}$ CCDW $\rho_{1Q^\mathrm{C}}(x)=\rho_0\cos\left( Q_1 x\right)$ (bottom of Fig. \ref{Fig4}(a)). This situation is clearly not energetically favorable as the CDW phase is stretched further and further along the boundary, paying a growing penalty from the derivative terms in the free energy [Eq. \eqref{eq:GL}].

Figs. \ref{Fig4}(c)-(d) show an alternative way for the system to accommodate the momentum difference $\bm \delta$ along the interface between $1Q^\mathrm{C}$ and $3Q^\mathrm{I}$ regions, by abruptly loosing a wavefront via a local $2\pi$ phase slip, i.e. in a lengthscale $\Delta x \ll 2\pi/\delta$. This mechanism is reminiscent of the theory of discommensurations introduced by McMillan to describe the phase transition between ICDWs and CCDWs as a series of phase slips \cite{McMillan76}, and happening here at the interface of those states instead. This local commensurability adjustment of the phase displays a vortex in its gradient $\bm{\nabla}\varphi(\bm r)$ [red arrows in Fig. \ref{Fig4}(c)] where a wavefront is locally lost, and it is the mechanism observed by STM at the boundaries between the $1Q^\mathrm{C}$ and $3Q^\mathrm{I}$ CDW regions in 4Ha [Fig. \ref{Fig3}(h)].

{ \sl Conclusion} --- We have shown that changing only the stacking sequence is sufficient to reshape both charge order and superconductivity in NbSe$_2$. The 4Ha polytype stabilizes a distinct electronic landscape, where competing charge-density-wave states coexist and superconductivity differs from that of the 2H phase. These findings demonstrate that polytypism provides a clean and intrinsic route to tune collective order in metallic van der Waals systems. Beyond NbSe$_2$, our results suggest that stacking engineering can be used as a general strategy to access, control, and potentially design emergent quantum phases in layered materials and more complex heterostructures.

{ \sl Acknowledgments} --- M.N.G. is supported by the Ramon y Cajal Fellowship RYC2021-031639-I funded by MCIN/AEI/10.13039/501100011033. M.M.U. and M.N.G. also acknowledge support from grant PID2023-153277NB-I00 from the same institution. M.M.U. acknowledges support by the ERC Starting grant LINKSPM (Grant \#758558). H.G. acknowledges funding from the EU NextGenerationEU/PRTR-C17.I1, as well as by the IKUR Strategy under the collaboration agreement between Ikerbasque Foundation and DIPC on behalf of the Department of Education of the Basque Government. S.M.V., C.B.C, and E.C. acknowledge financial support from the  Spanish Ministerio de Ciencia, Innovación y Universidades (MICIU) (Projects 2D-SPICE, ref. PID2023-149309OB-I00, cofinanced by FEDER; and Excellence Unit “María de Maeztu”, ref. CEX2024-001467-M; Ramón y Cajal program RYC2024-048264-I) and the Generalitat Valenciana (PROMETEO Program, CIPROM/2024/51). This study forms part of the Cátedra PERTE CHIP (ref.TSI-069100-2023-0012) and was supported by the Spanish Ministerio de Asuntos Económicos y Transformación Digital (A.Ec./T.D.) with funding from European Union NextGenerationEU. S.S. acknowledges enrollment in the doctorate program “Physics of Nanostructures and Advanced Materials” from the “Advanced polymers and materials, physics, chemistry and technology” department of the Universidad del País Vasco (UPV/EHU). F.J. acknowledges funding from a 2024 Leonardo Grant for Scientific Research and Cultural Creation from the BBVA Foundation. The work was also supported by the U.S. Department of Energy, Office of Basic Energy Sciences, Division of Materials Sciences and Engineering under Award DE-SC0025594 (development of local wave-vector measurement algorithm).

\bibliography{biblio} 
\section{ End Matter}
\subsection{Crystal Growth and Structural Characterization}

Single crystals of both 2H-NbSe$_2$ and 4Ha-NbSe$_2$ were synthesized using the chemical vapor transport (CVT) method with iodine (I$_2$) as the transport agent.

{ \sl 2H-NbSe$_2$ crystal growth}: Polycrystalline 2H-NbSe$_2$ precursor was first synthesized by mixing stoichiometric amounts of Nb (99.99\%, Alfa Aesar) and Se (99.999\%, Alfa Aesar). The mixture was sealed in an evacuated quartz ampoule ($P \sim 5 \times 10^{-5}$ mbar; length = 25 cm, inner diameter = 1.5 cm), heated to 910~$^\circ$C at a rate of 5~$^\circ$C/min, held for 9 days, and then cooled naturally to room temperature. Subsequently, 4 mmol of the obtained polycrystalline material was loaded into a second evacuated quartz ampoule together with iodine transport agent ([I$_2$] $\sim$ 5 mg/cm$^3$). The ampoule was placed in a three-zone furnace, heated to 700~$^\circ$C at 4~$^\circ$C/min, and maintained for 2 days. A temperature gradient of approximately 50~$^\circ$C was then established by increasing the source-zone temperature to 750~$^\circ$C. Crystal growth was continued for 15 days, followed by natural cooling to room temperature.

The obtained crystals were characterized by inductively coupled plasma (ICP) analysis and powder X-ray diffraction (XRD). The measured composition was Nb: $(36.0 \pm 1.0)\%$ and Se: $(62 \pm 2)\%$, consistent with the nominal stoichiometry of NbSe$_2$ (Nb: 37.0\%, Se: 63.0\%). Rietveld refinement confirmed the hexagonal 2H structure with space group $P6_3/mmc$, yielding lattice parameters $a=b=3.4450(3)$~\AA{} and $c=12.546(2)$~\AA{}, consistent with previous reports~\cite{Meerschaut2001}.

{ \sl 4Ha-NbSe$_2$ crystal growth}: Single crystals of 4Ha-NbSe$_2$ were also synthesized by the CVT method. Pre-reacted polycrystalline 4Ha-NbSe$_2$ powder was sealed under vacuum in a quartz ampoule together with a small amount of iodine (I$_2$) as the transport agent. The ampoule was placed in a horizontal tubular furnace under a controlled temperature gradient, with the hot and cold zones maintained at 950~$^\circ$C and 850~$^\circ$C, respectively. After a growth duration of 15 days, the ampoule was rapidly quenched in ice water, resulting in flat, shiny crystals with millimeter-scale lateral dimensions collected from the cold zone.

Room-temperature X-ray diffraction (XRD) measurements were performed on both single-crystal and powdered samples using a PANalytical diffractometer with Cu K$_\alpha$ radiation ($\lambda = 1.54056$~\AA{}). The diffraction pattern obtained from the flat crystal surface exhibited only $(00l)$ reflections, indicating high crystallinity and strong preferred orientation along the $c$-axis. Powder XRD refinement confirmed a hexagonal crystal structure with space group $P\bar{6}m2$ (No.~187), with refined lattice parameters $a=b=3.441(8)$~\AA{} and $c=25.226(1)$~\AA{}.
\subsection{STM/STS measurements}
Experiments were performed at the Donostia International Physics Center,  using a commercial scanning tunneling microscope (USM1300, Unisoku) operating under ultrahigh vacuum conditions, with base temperatures of 4.2~K and 0.34~K and magnetic fields up to 11~T applied perpendicular to the sample surface. All measurements were carried out in the temperature range of 0.34--4.7~K.

To ensure reliable data acquisition, Pt/Ir tips were pretreated on Au(111) or Cu(111) surfaces and calibrated using their respective Shockley surface states. Scanning tunneling spectroscopy (STS) measurements were acquired using a standard lock-in technique, in which a peak-to-peak a.c. modulation voltage ($V_{\mathrm{ac}}$) of typically 6--30~$\mu$V at a frequency of $f = 833$~Hz was added to the sample bias voltage during data acquisition. All STM/STS data were post-processed and analyzed using the WSxM software package~\cite{Horcas2007}.

\subsection{Ginzburg Landau theory}

The phase diagram in Fig. \ref{Fig3}(e) is obtained by minimizing the layer free energy in Eq.\eqref{eq:GL} with trial states for 1Q $\phi_n=(\phi,0,0)$ and 3Q $\phi_n=(\phi,\phi,\phi)$ states. In commensurate states $\bm{\partial} \phi_n=0$, while in incommensurate states $i\bm{\partial} \phi_n= \bm{\delta}_n$, where $\bm{\delta}_n=\delta\bm{G}_n/|\bm{G}_n|$ with $|\bm{\delta}_n|=\delta$ with units of momentum. Evaluating the free energies for the different cases we obtain   

\begin{align}
    \mathcal{F}^{1Q^I}_\phi=&  
    A\phi^2+D\phi^4+E \phi^7+F\phi^8 \\
    \mathcal{F}^{1Q^C}_\phi=&
    (A+A_\delta \delta^2)\phi^2+D\phi^4+F\phi^8 \\
    \mathcal{F}^{3Q^I}_\phi=&3A\phi^2 +B \phi^3+3(C+D)\phi^4 \nonumber \\
    &+3E\phi^7+3F\phi^8 
\end{align}
which are separately minimized numerically with parameters $A=-100$, $A_\delta = 1/|G_n|^2$, $C=210$, $D=100$, $E=-5.5$, $F=10$, and the lowest energy among the three is chosen as the ground state. The phase diagram is then represented as a function of the dimensionless incommensurability $\delta/|G_n|$.

Note that since the 4Ha structure is non-centrosymmetric, the absence of inversion also allows ${\rm Im}[\phi_1\phi_2\phi_3]$ and ${\rm Im}[\phi_n^7]$ terms in Eq.\eqref{eq:GL}, but this does not change the general features described here.

\subsection{Local wave vector measurement algorithm}
To extract phase and strain information from STM topography, we employed an algorithm developed at Yale University to measure the local wave vector. The approach relies on the fact that, for a plane wave
\begin{equation}
I_0 = e^{i\mathbf{Q}\cdot\mathbf{r}}
\end{equation}
the wave fronts are perpendicular (normal) to the wave vector $\mathbf{Q}$. We then modulate the image with two test plane waves, $e^{i\mathbf{p}_1\cdot\mathbf{r}}$ and $e^{i\mathbf{p}_2\cdot\mathbf{r}}$, i.e., we multiply these waves by the original image. This yields two modulated waves whose wave-front angles $\theta_1$ and $\theta_2$ satisfy
\begin{align}
\frac{Q_y + p_{1, y}}{Q_x + p_{1, x}} = \frac{\sin{\theta_1}}{\cos{\theta_1}} \label{Eq: Modulation x}\\
\frac{Q_y + p_{2, y}}{Q_x + p_{2, x}} = \frac{\sin{\theta_2}}{\cos{\theta_2}} \label{Eq: Modulation y}
\end{align}
where $\theta_1$ and $\theta_2$ are the wave-front angles of the modulated waves. Because $\theta_1$ and $\theta_2$ can be measured with high accuracy using the Hough transform, the wave vector $\mathbf{Q}$ can be obtained by solving Eqs.\,\eqref{Eq: Modulation x} and \eqref{Eq: Modulation y}. This method avoids the fast Fourier transform (FFT) entirely and therefore does not suffer from the reduced $\mathbf{q}$-space resolution that arises from a limited field of view.

Local wave-vector measurement is particularly useful for probing strain and CDW incommensurability. As discussed in the main text, strain can locally distort the lattice and thereby shift the Bragg vectors. Using our algorithm, we map the large spatial variation of the Bragg vector in 2H-NbS$_2$, providing direct evidence of strain. In addition, the subtle difference between the CDW wave vectors in 1Q and 3Q regions can also be resolved by locally measuring the CDW wave vector.

It is worth noting that another widely used method for local wave-vector analysis is the Lawler-Fujita algorithm (LFA). The key difference is that LFA measures the phase directly, whereas our method measures the phase gradient. For a distorted plane wave with a nonlinear phase term,
\begin{equation}
I_0(\mathbf{r}) = e^{i(\mathbf{Q}_0\cdot\mathbf{r} + \varphi(\mathbf{r}))}
\end{equation}
LFA first shifts the wave vector to the origin, producing an image containing only the phase factor $e^{i \phi(\mathbf{r})}$. By taking the argument of this complex field and unwrapping it, one obtains the phase function $\varphi(\mathbf{r})$, which can then be used to compute physical quantities such as strain.

In contrast, measuring the local wave vector directly is equivalent to measuring the phase gradient. In a small region centered at $\mathbf{r}_0$, the phase admits a first-order expansion,
\begin{equation}
I_0(\mathbf{r}) = e^{i[\mathbf{Q}_0\cdot\mathbf{r} + \phi(\mathbf{r}_0) + \nabla\varphi(\mathbf{r}_0)\cdot\mathbf{r}]}
\end{equation}
showing that the locally observed wave vector is
\begin{equation}
\mathbf{Q}(\mathbf{r}) = \mathbf{Q}_0 + \nabla\varphi(\mathbf{r})
\end{equation}

For the purposes of this paper, measuring the phase gradient is more appropriate because the presence of vortices makes the global phase function ill-defined. As a result, the Lawler-Fujita procedure can produce unavoidable branch cuts in the unwrapped phase. The phase gradient, however, remains well defined: along any path that crosses a branch cut, the phase modulo $2\pi$ is still smooth, and therefore its local gradient is well behaved.

\clearpage

\clearpage
\onecolumngrid

\begin{center}

\setlength{\parindent}{0pt}
\setlength{\parskip}{0.6em}

\renewcommand{\normalsize}{\fontsize{11.5}{14}\selectfont}
\normalsize
{\Large Supplementary Information for}\\[1.5em]

{\huge \bfseries Stacking-tuned superconductivity and competing charge-density-wave states in NbSe$_2$}\\[3em]

{\large
Sandra Sajan,$^{1,*}$
Xinze Yang,$^{2,3,*}$
Haojie Guo,$^{1}$
Tarushi Agarwal,$^{4}$
Samuel Ma\~nas-Valero,$^{5}$
Carla Boix-Constant,$^{5,6}$
Eugenio Coronado,$^{5}$
Fernando de Juan,$^{1,7}$
Eduardo H. da Silva Neto,$^{2,3,8}$
Ravi P. Singh,$^{4}$
Maria N. Gastiasoro,$^{1,\dagger}$
and Miguel M. Ugeda$^{1,7,9,\ddagger}$
}\\[1.5em]

{\itshape
$^{1}$Donostia International Physics Center, Donostia-San Sebasti\'an, Spain\\
$^{2}$Department of Physics, Yale University, New Haven, CT, USA\\
$^{3}$Energy Sciences Institute, Yale University, West Haven, CT, USA
$^{4}$Department of Physics, Indian Institute of Science Education and Research Bhopal, Bhopal, India\\
$^{5}$Institute of Molecular Science (ICMOL), University of Valencia, Paterna-Valencia, Spain\\
$^{6}$Department of Materials Science \& Metallurgy, University of Cambridge, Cambridge, United Kingdom\\
$^{7}$Ikerbasque, Basque Foundation for Science, Bilbao, Spain\\
$^{8}$Department of Applied Physics, Yale University, New Haven, CT, USA\\
$^{9}$Centro de F\'isica de Materiales CSIC-UPV/EHU, San Sebasti\'an, Spain
}\\[1.5em]

\end{center}

\vspace{2em}

\setcounter{figure}{0}
\renewcommand{\thefigure}{S\arabic{figure}}
\clearpage
\setcounter{section}{0}
\renewcommand{\thesection}{S\arabic{section}}
\fontsize{12pt}{15pt}\selectfont

\section{Supplementary Note 1: Zoomed in superconducing gap}%
\vspace{4em}
\begin{figure}[ht]
    \includegraphics[width=0.35\textwidth]{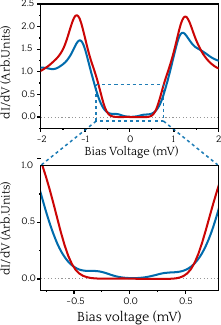}
    \caption{Comparison of the superconducting gap for both polytypes, with the zoom of the smaller gap of the 4Ha and 2H polytypes.}
    \label{fig:S1B}
\end{figure}

\section{Supplementary Note 2: Anisotropic Two-Band McMillan Analysis of Superconducting Tunneling Spectra}%
\label{sec:S1}
\begin{figure}[ht]
    \includegraphics[width=\columnwidth]{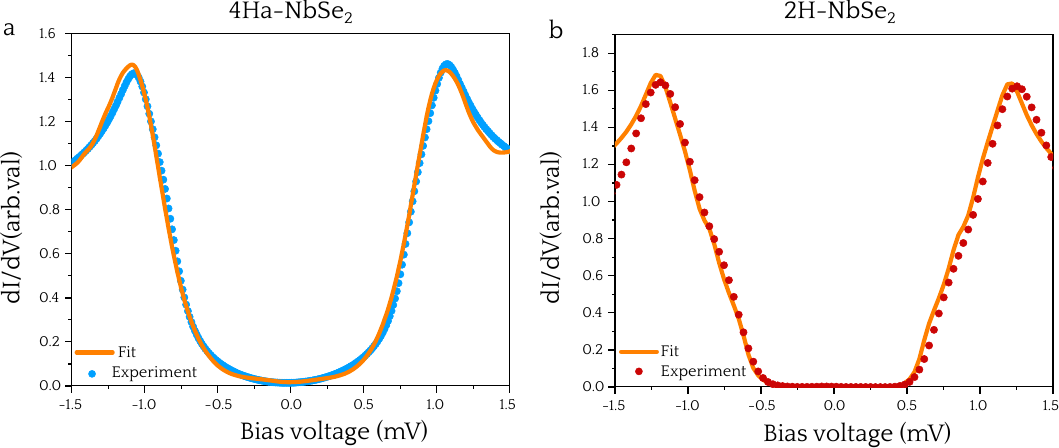}
   \caption{Comparison of scanning tunnelling spectroscopy (STS) spectra of (a)4Ha and (b)2H phases of NbSe$_2$ fitted using a two-band McMillan superconducting model.}
    \label{fig:S3}
\end{figure}

The superconducting tunneling spectra were analyzed using a self-consistent anisotropic two-band McMillan model~\cite{McMillan1968,Szabo2001}, which describes coupled multiband superconductivity through interband pairing interactions between two superconducting condensates. In this framework, the quasiparticle density of states is determined by the superconducting gap magnitudes ($\Delta_1$, $\Delta_2$), quasiparticle lifetime broadening parameters ($\Gamma_1$, $\Gamma_2$), interband coupling strengths ($\Gamma_{12}$, $\Gamma_{21}$), relative spectral weight ($w$), and anisotropic gap modulation parameters ($A_1$, $A_2$), which account for the sixfold in-plane symmetry of NbSe$_2$. The total tunneling density of states is written as
$N(E)=wN_1(E)+(1-w)N_2(E)$, where $N_1$ and $N_2$ are the band-resolved superconducting densities of states; thus, $w$ is the normalized tunneling weight of the first component. A linear background correction characterized by slope $a$ and offset $b$ was additionally included to account for weak spectral asymmetry and experimental background contributions.

The measured $dI/dV$ spectra of 4Ha-NbSe$_2$ exhibit a pronounced two-step superconducting structure consisting of coherence peaks near $\pm1.1~\mathrm{mV}$ and weaker low-energy shoulders near $\pm0.3~\mathrm{mV}$, corresponding to approximate peak-to-peak separations of 2.2~mV and 0.6~mV, respectively. These features qualitatively indicate the presence of multiple superconducting energy scales. Quantitative extraction of the superconducting parameters was obtained through nonlinear least-squares fitting using the anisotropic McMillan formalism.

Averaging over multiple spectra acquired on 4Ha-NbSe$_2$ at $T=340~\mathrm{mK}$ yields superconducting gap values of $\langle \Delta_1 \rangle = 0.987\pm 0.082~\mathrm{meV}$ and $\langle \Delta_2 \rangle = 0.269\pm 0.130~\mathrm{meV}$. The corresponding quasiparticle broadening parameters are $\langle \Gamma_1 \rangle = 0.142\pm0.053~\mathrm{meV}$ and $\langle \Gamma_2 \rangle = 0.160\pm 0.203~\mathrm{meV}$, while the interband coupling strengths are found to be $\langle \Gamma_{12} \rangle = 0.079\pm 0.140~\mathrm{meV}$ and $\langle \Gamma_{21} \rangle = 0.388\pm 0.397~\mathrm{meV}$. The extracted anisotropy parameters are moderate, with $\langle A_1 \rangle = 0.143\pm 0.057$ and $\langle A_2 \rangle = 0.174\pm0.368$, consistent with anisotropic multiband superconductivity expected in layered transition-metal dichalcogenides. The average spectral weight remains strongly dominated by the larger-gap component ($\langle w \rangle \approx 0.981\pm0.023$), suggesting that the dominant contribution to the tunneling conductance originates from the larger-gap Fermi-surface sheet.

In contrast, fitting of the 2H-NbSe$_2$ spectra yields significantly larger and better-resolved superconducting gaps, with $\Delta_1 = 1.08~\mathrm{meV}$ and $\Delta_2 = 0.72~\mathrm{meV}$. The corresponding quasiparticle broadenings are substantially smaller, $\Gamma_1 = 0.04~\mathrm{meV}$ and $\Gamma_2 = 0.05~\mathrm{meV}$, indicating a reduced quasiparticle scattering. The interband coupling parameters obtained for 2H-NbSe$_2$ are $\Gamma_{12} = 0.097~\mathrm{meV}$ and $\Gamma_{21} \approx 0$, suggesting comparatively weak asymmetric interband coupling. The extracted anisotropy parameters, $A_1 = 0.116$ and $A_2 = 0.167$, remain moderate and are consistent with anisotropic superconductivity associated with the hexagonal crystal symmetry of NbSe$_2$. Unlike the 4Ha phase, the relative spectral weight in 2H-NbSe$_2$ is distributed between both superconducting bands ($w \approx 0.68$), indicating a more substantial contribution from the smaller-gap band to the tunneling conductance.

The low-energy tunneling spectra further reveal qualitative differences between the two polytypes. In 2H-NbSe$_2$, the superconducting gap exhibits a flatter low-bias conductance and sharper coherence peaks, characteristic of a cleaner and more homogeneous superconducting condensate. By contrast, the 4Ha spectra display broader peaks and a more rounded low-energy density of states, reflecting enhanced quasiparticle scattering and reduced visibility of the smaller superconducting gap. In several 4Ha spectra, the smaller gap appears only as a weak shoulder near the Fermi level rather than as a distinct coherence peak. This behavior likely originates from stacking-induced modifications of the electronic structure, including altered interlayer hybridization and selective suppression of superconductivity on one Fermi-surface sheet.

Magnetization and transport experiments establish superconducting transition temperatures of $T_c \approx 6.2~\mathrm{K}$ for 4Ha-NbSe$_2$ and $T_c \approx 7.2~\mathrm{K}$ for 2H-NbSe$_2$, together with charge-density-wave transition temperatures of $T_{\mathrm{CDW}} \approx 41~\mathrm{K}$ and $T_{\mathrm{CDW}} \approx 33~\mathrm{K}$, respectively~\cite{Patra2025,Yokoya2001,Moncton}. Using $T_c = 6.2~\mathrm{K}$, the larger superconducting gap in 4Ha-NbSe$_2$ yields a ratio of $2\Delta_1/k_B T_c \approx 3.7$, close to the weak-coupling BCS value of 3.53. The corresponding ratio for 2H-NbSe$_2$ is larger, consistent with stronger multiband superconductivity and enhanced condensate coherence. Overall, these results demonstrate that multiband superconductivity persists in both structural polytypes; however, the secondary superconducting gap is substantially weakened in 4Ha-NbSe$_2$, suggesting that the modified stacking sequence selectively suppresses superconductivity on one portion of the Fermi surface while preserving the dominant superconducting condensate.

\section{Supplementary Note 3: Magnetic field dependence}%
\begin{figure}[ht]
    \includegraphics[width=0.8\columnwidth]{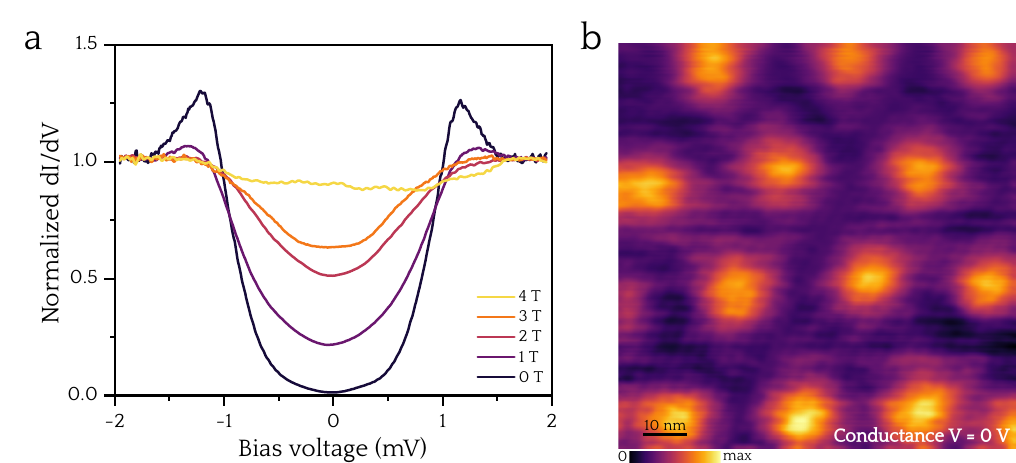}
    \caption{(a) Representative $dI/dV$ spectra showing the evolution of the superconducting gap with magnetic field on the 4Ha-NbSe$_2$ polytype. 
   (b) Vortex lattice mapping at a magnetic field of 2.5T ($V_s = 0\mathrm{V}$). All measurements were performed at $T = 340\,\mathrm{mK}$.}
    \label{fig:S1}
\end{figure}

We discuss the magnetic-field dependence of the superconducting gap in the 4Ha polytype of NbSe$_2$, showing that the larger superconducting gap is progressively suppressed with increasing magnetic field and vanishes at approximately 4~T, corresponding to an upper critical field of $B_{c2} \approx 4$~T.To support this result, we provide a derivation based on vortex-lattice imaging and coherence length estimates. Scanning probe measurements at 2.5~T reveal a well-ordered Abrikosov vortex lattice with a periodicity of $a \approx 30$~nm. The vortex cores have an estimated diameter of approximately 14--18~nm, yielding a superconducting coherence length of $\xi \approx 7$--9~nm.Using the standard relation between coherence length and upper critical field\cite{Tinkham2004},
\[
B_{c2} = \frac{\Phi_0}{2\pi \xi^2},
\]
we estimate an upper critical field in the range of 4--6~T, consistent with both the vortex-core size analysis and the experimentally observed suppression of the superconducting gap.

\clearpage
\clearpage
\section{Supplementary Note 4: Comparing the $\nabla\varphi$ of the Bragg vector and CDW }
Figure 3g shows vortices at the boundary separating the 1Q and 3Q regions. To demonstrate that these vortices are not induced by lattice dislocations, we provide the corresponding vector-field plot of $\delta\mathbf{Q}$ extracted from the Bragg peak, which is clearly vortex-free.

\begin{figure}[h]
    \centering
    \includegraphics[width=0.9\linewidth]{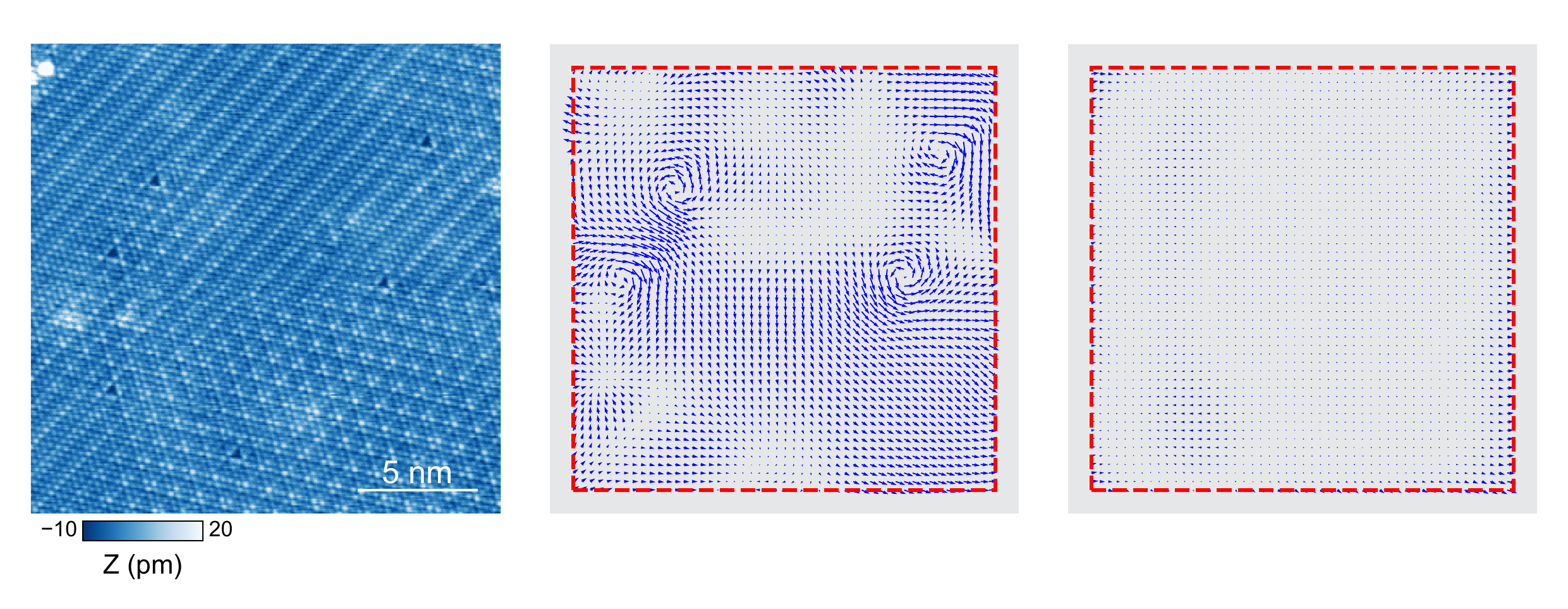}
    \caption{(a) Topography of the region exhibiting coexistence of 1Q and 3Q CDW. (b) Vector field of $\nabla \varphi_{\mathrm{CDW}}$. (c) Vector field of $\nabla \varphi_{\mathrm{Bragg}}$.}
    \label{fig:figureS_Data059}
\end{figure}

\end{document}